\documentclass[]{spie}  
\usepackage[]{amsmath}
\usepackage[]{amssymb}
\usepackage[]{graphicx}
\usepackage{lmodern}

\usepackage{xspace} 

\def\eg{{\it e.g.}\xspace}
\def\etal{{\it et al.}\xspace}
\def\spider{{\sc Spider}\xspace}
\def\boomerang{{\sc BOOMERanG}\xspace}
\def\blast{{\sc BLAST}\xspace}
\def\ebex{{\sc EBEX}\xspace}
\def\planck{{\it Planck}\xspace}
\def\bicep{{\sc Bicep}\xspace}
\def\mukcmbrts{$\mu$K$_{cmb}$$\sqrt{s}$}

\title{\spider: a balloon-borne CMB \\
polarimeter for large angular scales} 

\author{J.P.~Filippini\supit{a}*, P.A.R.~Ade\supit{b},
M.~Amiri\supit{c}, S.J.~Benton\supit{d},
R.~Bihary\supit{e}, J.J.~Bock\supit{a,f},
J.R.~Bond\supit{g},\\ J.A.~Bonetti\supit{f},
S.A.~Bryan\supit{e}, B.~Burger\supit{c}, H.C.~Chiang\supit{h}, 
C.R.~Contaldi\supit{i}, B.P.~Crill\supit{a,f}, 
O.~Dor\'{e}\supit{a,f}, M.~Farhang\supit{d},
L.M.~Fissel\supit{d}, N.N.~Gandilo\supit{d},
S.R.~Golwala\supit{a}, J.E.~Gudmundsson\supit{h},
M.~Halpern\supit{c}, M.~Hasselfield\supit{c}, 
G.~Hilton\supit{j}, W.~Holmes\supit{f},
V.V.~Hristov\supit{a}, K.D.~Irwin\supit{j},
W.C.~Jones\supit{h}, C.L.~Kuo\supit{k},
C.J.~MacTavish\supit{l}, P.V.~Mason\supit{a}, 
T.E.~Montroy\supit{e}, T.A.~Morford\supit{a}, 
C.B.~Netterfield\supit{d}, D.T.~O'Dea\supit{i}, 
A.S.~Rahlin\supit{h}, C.D.~Reintsema\supit{j},
J.E.~Ruhl\supit{e}, M.C.~Runyan\supit{a},
M.A.~Schenker\supit{a}, J.A.~Shariff\supit{d},
J.D.~Soler\supit{d}, A.~Trangsrud\supit{a},
C.~Tucker\supit{b}, R.S.~Tucker\supit{a}, and
A.D.~Turner\supit{f}
\skiplinehalf
\supit{a}Division of Physics, Mathematics, and Astronomy, California Institute of Technology,
Pasadena, CA, USA; \\
\supit{b}School of Physics and Astronomy, Cardiff University, Cardiff, UK; \\
\supit{c}Department of Physics and Astronomy, University of British
Columbia, Vancouver, BC, Canada; \\
\supit{d}Department of Physics, University of Toronto, Toronto, ON,
Canada; \\
\supit{e}Department of Physics, Case Western Reserve University,
Cleveland, OH, USA; \\
\supit{f}Jet Propulsion Laboratory, Pasadena, CA, USA; \\
\supit{g}Canadian Institute for Theoretical Astrophysics, University
of Toronto, Toronto, ON, Canada; \\
\supit{h}Department of Physics, Princeton University, Princeton, NJ, USA; \\
\supit{i}Department of Physics, Imperial College, University of
London, London, UK; \\
\supit{j}National Institute of Standards and Technology, Boulder, CO, USA; \\
\supit{k}Department of Physics, Stanford University, Stanford, CA, USA; \\
\supit{l}Kavli Institute for Cosmology, University of Cambridge, Cambridge, UK }

\authorinfo{*Corresponding author; E-mail: jpf@caltech.edu, Tel: 1 626 395 2601}
 
\begin{document} 
  \maketitle 

\begin{abstract}
We describe \spider, a balloon-borne instrument to map the polarization of the millimeter-wave sky with degree angular resolution.  \spider consists of six monochromatic refracting telescopes, each illuminating a focal plane of large-format antenna-coupled bolometer arrays.  A total of 2,624 superconducting transition-edge sensors are distributed among three observing bands centered at 90, 150, and 280 GHz.  A cold half-wave plate at the aperture of each telescope modulates the polarization of incoming light to control systematics.  \spider's first flight will be a 20-30-day Antarctic balloon campaign in December 2011.  This flight will map $\sim$8\% of the sky to achieve unprecedented sensitivity to the polarization signature of the gravitational wave background predicted by inflationary cosmology.  The \spider mission will also serve as a proving ground for these detector technologies in preparation for a future satellite mission.
\end{abstract}





\section{INTRODUCTION}
\label{sec:intro}  

Over the past 45 years, observations of the cosmic microwave background (CMB) have transformed our understanding of the universe.  The discovery of this primordial radiation field\cite{Penzias:1965wn} heralded the triumph of the hot Big Bang theory, laying down a timeline for cosmological history.  The discovery of tiny (one part in $10^5$) fluctuations in the temperature of this radiation across the sky\cite{Smoot:1992td} gave cosmologists a snapshot of the birth of large-scale structure in our universe, as well as evidence that its history began with an inflationary epoch of rapid expansion.  More recently, a series of terrestrial, balloon-borne, and satellite instruments have taken advantage of rapid advances in millimeter-wave technology to map these anisotropies with ever greater fidelity.  These measurements continue to provide an ever-more-precise picture of the contents, geometry, and history of the cosmos\cite{Komatsu:2010fb}.

The next frontier of this endeavor is the measurement of the tiny degree of polarization in the microwave background radiation.  Thomson scattering at the epoch of recombination transformed quadrupole anisotropies in the photon-baryon fluid into a slight polarization in the CMB, with an intensity more than an order of magnitude smaller than that of the temperature anisotropies.  The pattern of this polarization across the sky encodes information about the local conditions within the last-scattering surface that is complementary to that available in the temperature anisotropies.  Early measurements of CMB polarization have already provided independent confirmation of our understanding of recombination and improved our estimates of cosmological parameters\cite{Brown:2009uy,Chiang:2009xsa,Komatsu:2010fb}.  

In addition to improved constraints on existing cosmological parameters, future precision measurements of CMB polarization promise qualitatively new information about the history of our universe.  The large-scale polarization pattern encodes unique data on two epochs of cosmological history that have thus far been hidden from observation: the reionization of the universe by the ignition of the first stars, and the inflationary period of rapid expansion in the first $\sim$$10^{-33}$ second after the Big Bang.  Reionization generates new opportunities for Thomson scattering of CMB photons, and thus generates additional polarization of the microwave sky.  The history of reionization is thus encoded in the CMB polarization pattern on the very largest angular scales, easily distinguished from the signatures of recombination at smaller angular scales.  Gravitational waves generated during the inflationary era induce polarization during the recombination and reionization epochs.  These gravitational waves are unique in their ability to generate a ``B-mode'' polarization pattern at large angular scales, distinguishable from the more common ``E-mode'' pattern by its opposite transformation under spatial inversion \cite{Kamionkowski:1996ks}.  The detection of a B-mode pattern on large angular scales would be a dramatic confirmation of the inflationary theory of cosmology.  The CMB polarization pattern also holds valuable information about more recent cosmological history, encoded in the distortion of the primordial polarization map by gravitational lensing.

In this paper we describe \spider, a balloon-borne telescope array optimized to search for the signature of cosmic inflation in the polarization of the CMB.  \spider will take advantage of massively multiplexed bolometer arrays and a vantage point above the bulk of the atmosphere to map the polarization of the millimeter-wave sky with higher fidelity than any CMB receiver to date.  The instrument and its scan strategy are optimized to map a large fraction of the sky at degree angular scales, an ideal approach for a targeted search for the ``B-mode'' signature of inflation.  \spider will use stepped half-wave plates and broad frequency coverage to achieve excellent control of instrumental systematics and contamination from galactic foregrounds.  Following in the tradition of \boomerang and \blast, \spider will also serve as a technology pathfinder for a satellite mission to make definitive maps of CMB polarization.

The \spider instrument has previously been described by Montroy \etal\cite{Montroy:2006spie} and Crill \etal\cite{Crill:2008spie} in this conference series.  MacTavish \etal\cite{MacTavish:2008syst} have presented a detailed study of instrumental systematics for \spider, focusing on the case of a 2-6 day flight from Alice Springs, Australia.  In this work we describe the present status of \spider, which is currently targeting an Antarctic flight during the 2011-12 austral summer.  Three other contributions to these proceedings describe particular subsystems of \spider in greater detail: Bryan \etal\cite{Bryan:2010spie}, Gudmundsson \etal\cite{Gudmundsson:2010spie}, and Runyan \etal\cite{Runyan:2010spie}.

\section{SCIENTIFIC GOALS}
\label{sec:science}  

The primary goal of the \spider project is to search for the signature of a primordial gravitational wave background in the polarization of the cosmic microwave background, a key prediction of the inflationary paradigm of cosmogenesis.  To this end, \spider will map the polarization of the millimeter sky with high fidelity on the largest scales allowed by the presence of Galactic emission.  The \spider polarization map will also address two secondary science goals, also of great scientific importance: measurement of the weak lensing of the CMB by cosmological structure and characterization of polarized emission from interstellar dust at high galactic latitudes.  The raw sensitivity of its receivers and the excellent control of systematics afforded by its instrument design will give \spider's data set lasting value.

\begin{figure}[tbph]
\begin{center}
\includegraphics[height=2.7in]{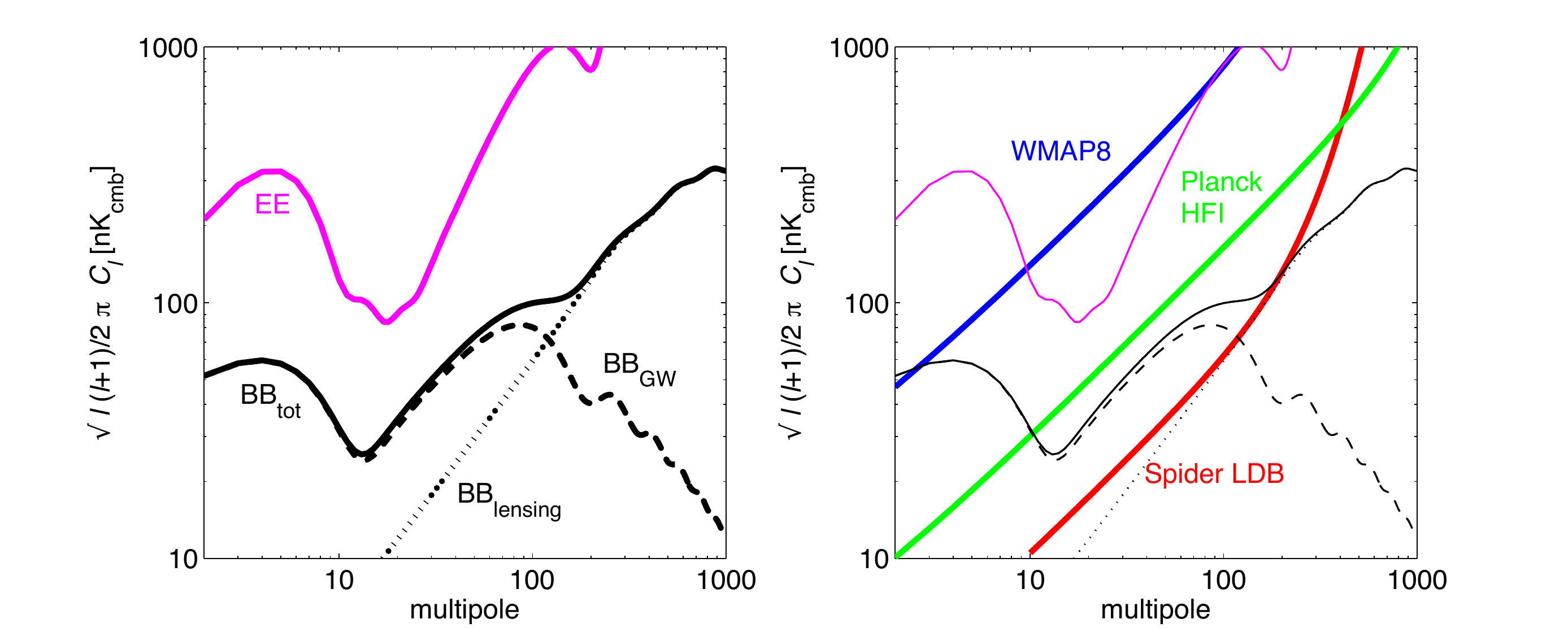} \\
\caption{\small An illustration of the B-mode angular power spectra and the noise
levels associated with the WMAP 8-year, \planck, and \spider surveys.
The left panel shows the theoretical spectra for the E-modes (labelled EE),
gravitational lensing B-modes (labelled BB$_{lensing}$), and inflationary
gravitational wave B-modes (labelled BB$_{GW}$) for $r = 0.1$.
The right panel overlays the noise thresholds per multipole
(including both statistical noise and sample variance) for the
three surveys.  
The curves for WMAP and \planck are derived from a simple Fisher analysis.  For
\spider, the Fisher errors are corrected for instrumental effects using
ensembles of time domain simulations that include the scan strategy, noise
correlations, and stationarity boundaries. The line for \spider has been truncated at a multipole of ten
as the limited sky coverage of an Antarctic flight will prevent us from probing larger scales.  The effects of 
foreground emission are not included in these results; in particular, it is
possible that foreground emission will limit \planck's measurement 
of the low-$\ell$ BB bump induced by reionization.  
}
\label{fig:cl_spectra}
\end{center}
\end{figure}

\begin{figure}[tbph]
\begin{center}
\includegraphics[height=2.8in]{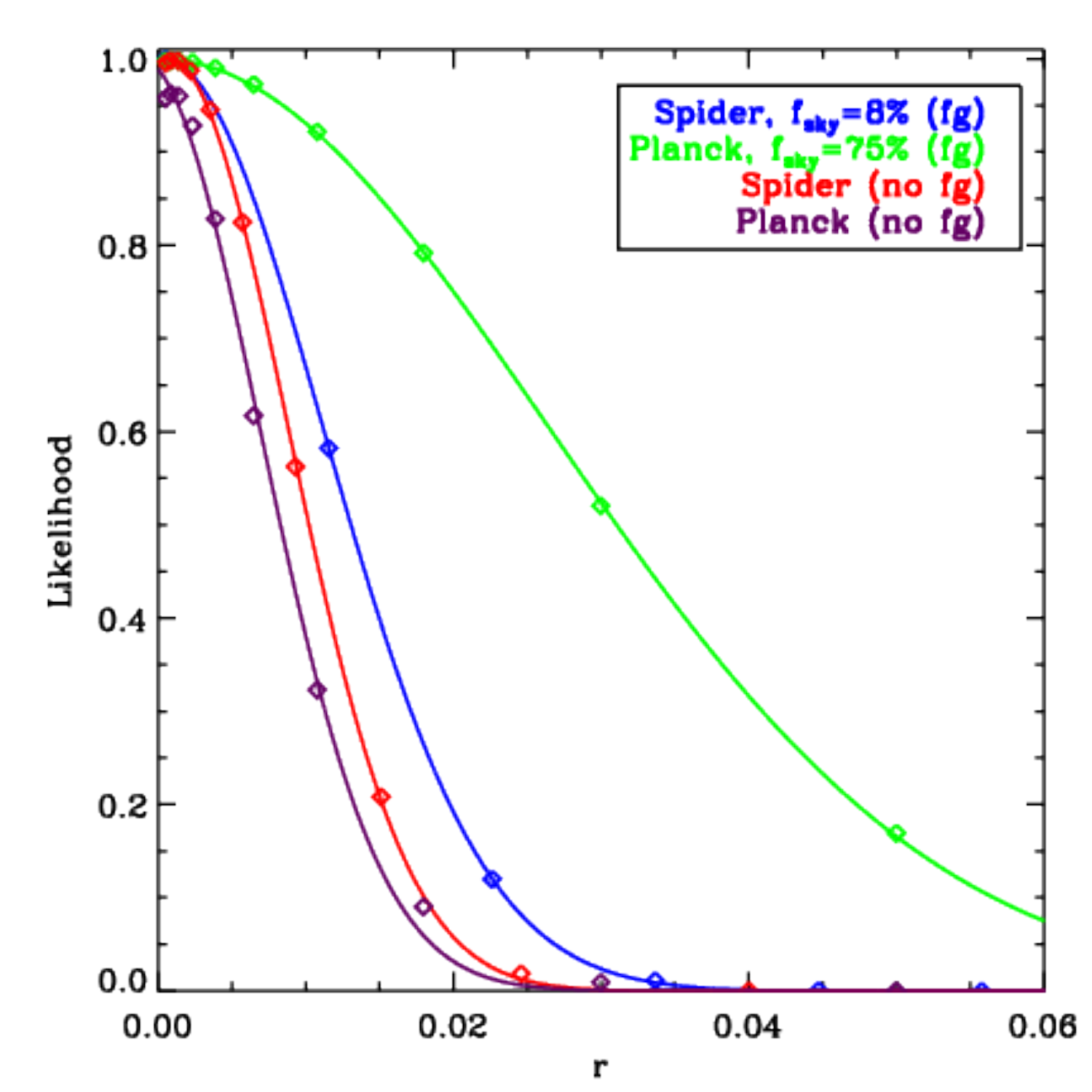}
\includegraphics[height=2.8in]{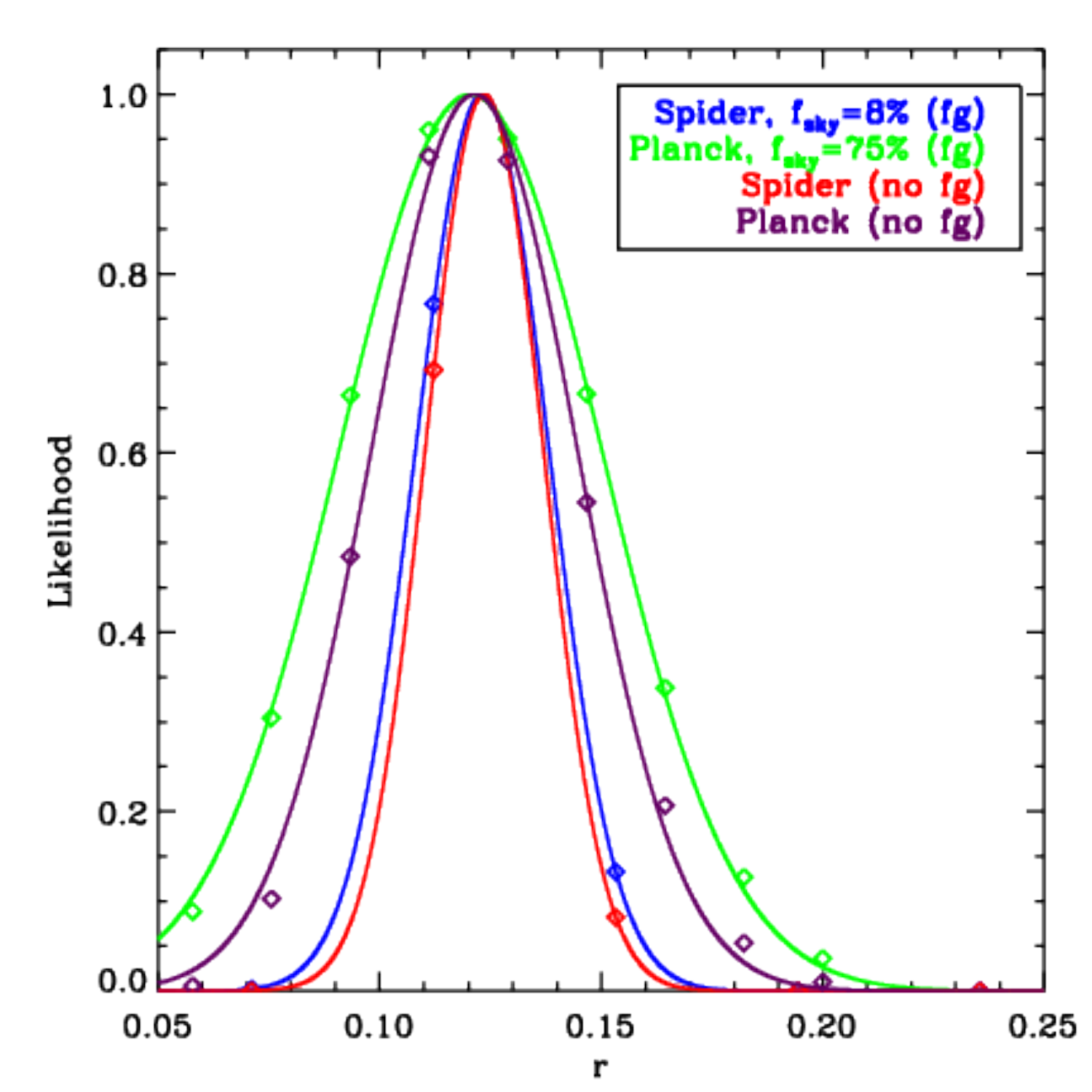}
\end{center}
\caption{\small \emph{At left:} the tensor to scalar likelihood, in the limit
of a small tensor to scalar ratio ($r=0.001$, for \spider and \planck
with and without including the impact of foregrounds.
\emph{At right:} the same, for the case of a relatively high tensor to scalar
ratio, $r=0.12$, corresponding to a simple slow roll inflation ($m^2\phi^2$).
  Both assume an optical depth to reionization of $\tau = 0.09$.
\planck's ability to measure or place limits on $r$ is due entirely to the
B-mode signature coming from the epoch of reionization on angular scales
larger than $\sim 10^\circ$, where Galactic foregrounds are dominant.
The dramatic impact of these foregrounds on \planck's $r$ measurement are
evident.
}\label{fig:r_marg}
\end{figure}

\subsection{The echoes of inflation}
Models of inflation\cite{Guth:1980zm} postulate that the universe underwent a period of extremely rapid expansion during its first moments, increasing in scale by a factor of $\gtrsim$$10^{25}$ in a mere $\sim$$10^{-33}$ second.  This massive expansion left the universe nearly spatially-flat and remarkably homogeneous and isotropic on its largest scales.  Quantum fluctuations in the ``inflaton'' field (or fields) that caused this expansion seeded tiny inhomogeneities into the universe's energy density.  Over time these ripples grew under their own gravity, eventually evolving into the large-scale structures of the modern universe.  Though this basic picture has passed every observational test to date, the identity of the inflaton field that started it all remains unknown.  The particle physics behind inflation is likely to occur at energy scales far above those accessible to particle accelerators, so cosmological measurements are our primary avenue for investigating inflationary physics.

A unique prediction of inflation is that these quantum fluctuations should have seeded the early universe with both scalar and tensor perturbations of the underlying spatial metric over a wide range of distance scales.  The former grew into the large-scale structure we observe today; the latter manifested themselves as a stochastic background of gravitational waves.  The long wavelengths of these primordial gravitational waves render them essentially undetectable today (at least with near-term technology), but their presence at the epoch of recombination would have imparted a unique signature to the polarization of the CMB: a B-mode contribution peaking at angular scales of a few degrees ($\ell\approx80$) \cite{Polnarev:1985,Crittenden:1993ni,Kamionkowski:1996zd}.  Both tensor and scalar perturbations generate polarization through the same mechanism (Thomson scattering), but symmetry requires that scalar perturbations generate only E-modes, while tensor perturbations generate a mixture of both.  Though gravitational lensing can convert E-modes to B-modes on sub-degree angular scales, no process in nature is expected to produce a B-mode signal on larger scales.  A detection of a B-mode component on large angular scales would thus be a clear confirmation that our universe began with an inflationary epoch.

The power in the inflationary gravitational waves relative to that in the usual scalar anisotropies is characterized by the tensor-to-scalar ratio, $r$.  Current cosmological data constrain $r$ relatively weakly: $r\lesssim0.20$ at the 95\% confidence level, with some dependence on cosmological assumptions\cite{Komatsu:2010fb}.  The value of $r$ is a function of the energy scale of the new physical processes behind inflation ($E_{inf}$), a scale which is not well constrained by theory.  {\it A priori}, $E_{inf}$ could lie anywhere between the TeV scale of current accelerator data and the Planck scale ($10^{19}$ GeV), and over most of this range the inflationary B-mode signal will be undetectable amidst the background from gravitational lensing of E-modes.  If inflation is associated with the physics of Grand Unified Theories (GUTs), however, its B-mode signature may be detectable: $r\gtrsim0.01$ corresponds to $E_{inf}\gtrsim10^{16}$ GeV.  Recent evidence that the spectral index of the primordial power spectrum is less than unity ($n_s=0.96\pm0.1$ as of this writing\cite{Komatsu:2010fb}) provides tantalizing support for this hypothesis: in many of the simplest single-field inflation models, a low spectral index corresponds to detectably large values of $r$\cite{Baumann:2009cmbpol}.  If an inflationary B-mode patten is detected, it will provide invaluable information about the nature of inflation and the fundamental physics responsible for it.

Figure \ref{fig:cl_spectra} shows theoretical power spectra for the E-mode and B-mode polarization components, under the assumption that $r=0.1$.  The inflationary contribution to the B-mode spectrum peaks at angular scales of a few degrees ($\ell$$\sim$80).  At smaller scales this signal is contaminated by the gravitational lensing of E-modes; at the largest scales one cannot avoid galactic foregrounds.  This suggests a clear strategy for a targeted measurement of $r$: a suborbital mission to map the CMB polarization with extremely high fidelity on relatively large angular scales ($10\lesssim\ell\lesssim300$).  \spider is designed to implement this strategy, combining simple degree-scale optics with the low foregrounds and large sky coverage afforded by Antarctic ballooning.  This strategy is extremely complementary to that of \planck, which achieves its sensitivity to $r$ solely from the lowest-$\ell$ reionization peak in the B-mode spectrum.  Figure \ref{fig:r_marg} compares the predicted effect of galactic foregrounds on each experiment's sensitivity to $r$.  Though the two instruments have comparable raw sensitivities to $r$, \spider's reach is hampered far less by the introduction of galactic foregrounds.  \spider is designed to measure the tensor-to-scalar ratio corresponding to single-field inflationary models near the GUT scale, or to experimentally rule them out by setting an interesting upper limit $r<0.03$ (at 3$\sigma$).

\subsection{Weak lensing of the microwave background}
As the CMB radiation propagates from the surface of last scattering to our telescope, it is lensed by the gravitational potential it traverses.  Such lensing distorts the polarization pattern we observe in the sky, converting a small portion of the E-mode power into a B-mode signal.  Though this lensing B-mode (shown in Figure \ref{fig:cl_spectra}) constitutes a limiting background to searches for a primordial B-signal, it is also a valuable probe of cosmology in its own right.  Through lensing, the B-mode polarization of the CMB is a sensitive probe of the matter power spectrum, and thus on such parameters as the dark energy equation and the sum of the neutrino masses \cite{dePutter:2009kn,Smith:2006nk}.  The great depth of \spider's polarization maps will allow it to probe the lensing of the CMB polarization field, with an anticipated signal-to-noise per multipole on the lensing signal (\eg for $r=0$) of order unity from $10 < \ell < 300$ (see Figure \ref{fig:cl_spectra}).

\subsection{Galactic dust emission}
Confusion with galactic foregrounds is likely to set the ultimate limit on measurements of CMB polarization on large angular scales.  WMAP has mapped the polarization of galactic synchrotron emission\cite{Gold:2010fm}, the dominant CMB foreground below $\sim$100 GHz, but the polarized contribution of dust emission at higher frequencies is not well constrained.  \spider will map the polarized emission of dust at high galactic latitude, providing valuable insight into the denser portions of the interstellar medium.  \spider's high-fidelity polarized dust maps will complement the all-sky maps produced by the \planck satellite, which will provide a precise determination of the overall intensity of dust emission.  The 280 GHz band of the \spider instrument is very complementary to \planck's frequency coverage, bridging the gap between the latter's bands at 217 GHz and and 353 GHz.

\section{MISSION}
\label{sec:mission}  

\spider's maiden voyage will be a long-duration balloon flight from McMurdo station, Antarctica, during the 2011-12 austral summer.  Due to the persistent daylight the instrument will scan in azimuth, with an amplitude of 45 degrees and a maximum azimuthal velocity of 6 degrees/second.  The scan elevation will be changed by 0.5 degrees each hour, covering the range between 28 and 40 degrees each day.  The half-wave plate will be operated in a stepped mode, turning by a discrete number of degrees several times each day.  The sky coverage of this strategy is enhanced by \spider's small, easily-shielded aperture, which allows it to scan closer to the sun than previous suborbital instruments such as \boomerang or \ebex.

\begin{figure}[!t]
\begin{center}
\scalebox{0.28}{\includegraphics[bb=-50 -20 700 420]{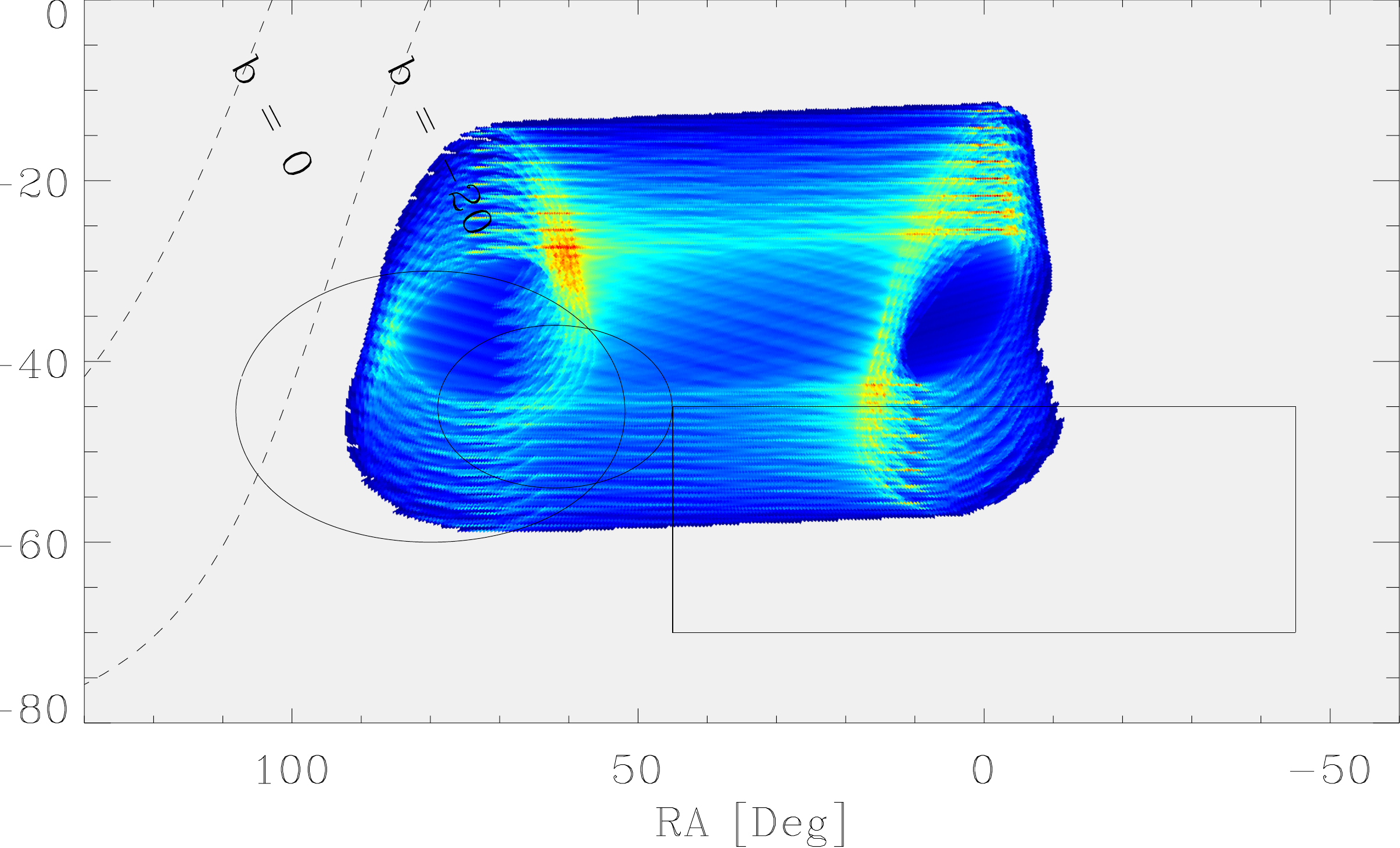}}
\scalebox{0.28}{\includegraphics[bb=-50 -20 700 420]{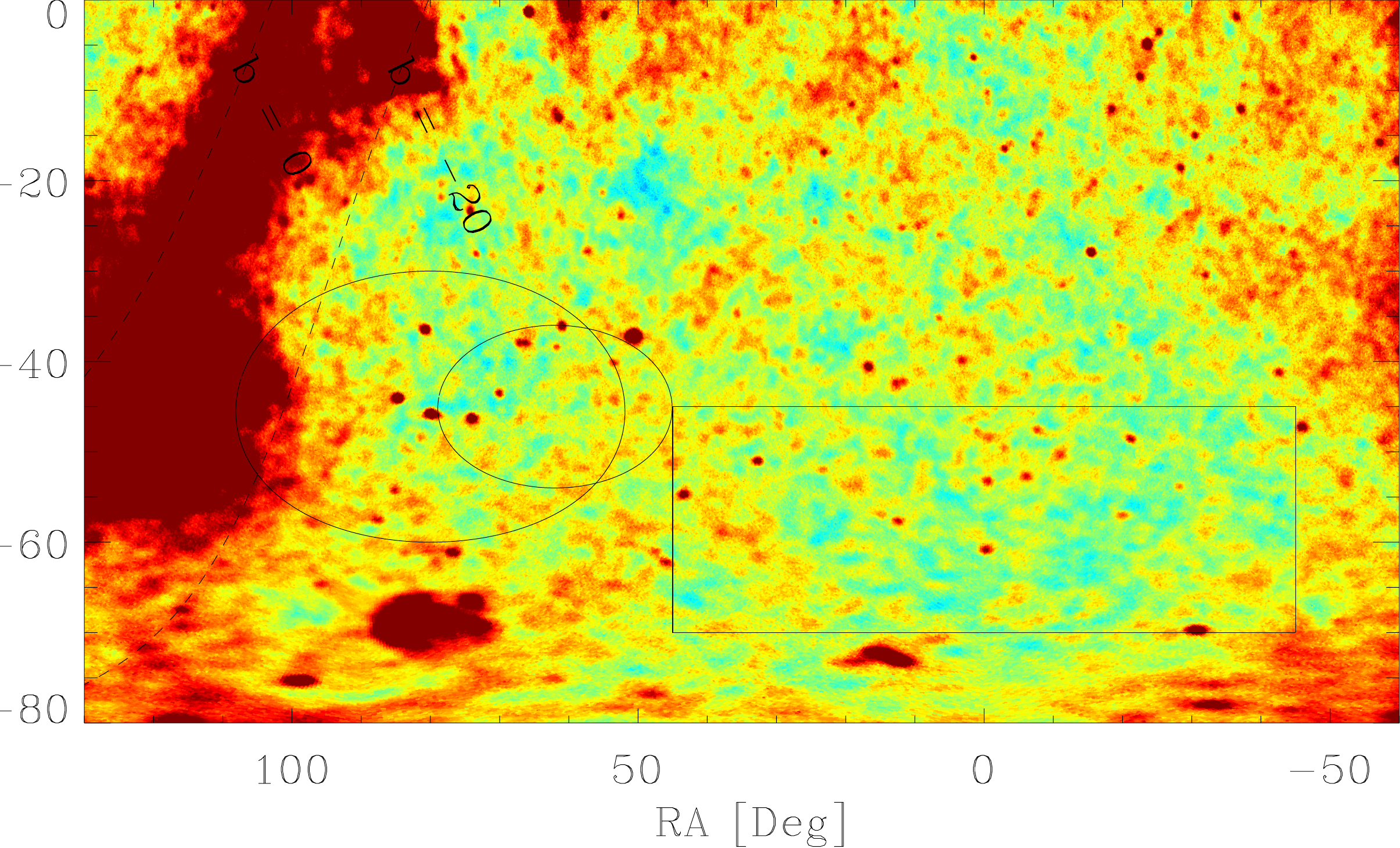}}

\scalebox{0.28}{\includegraphics[bb=-50 -20 700 420]{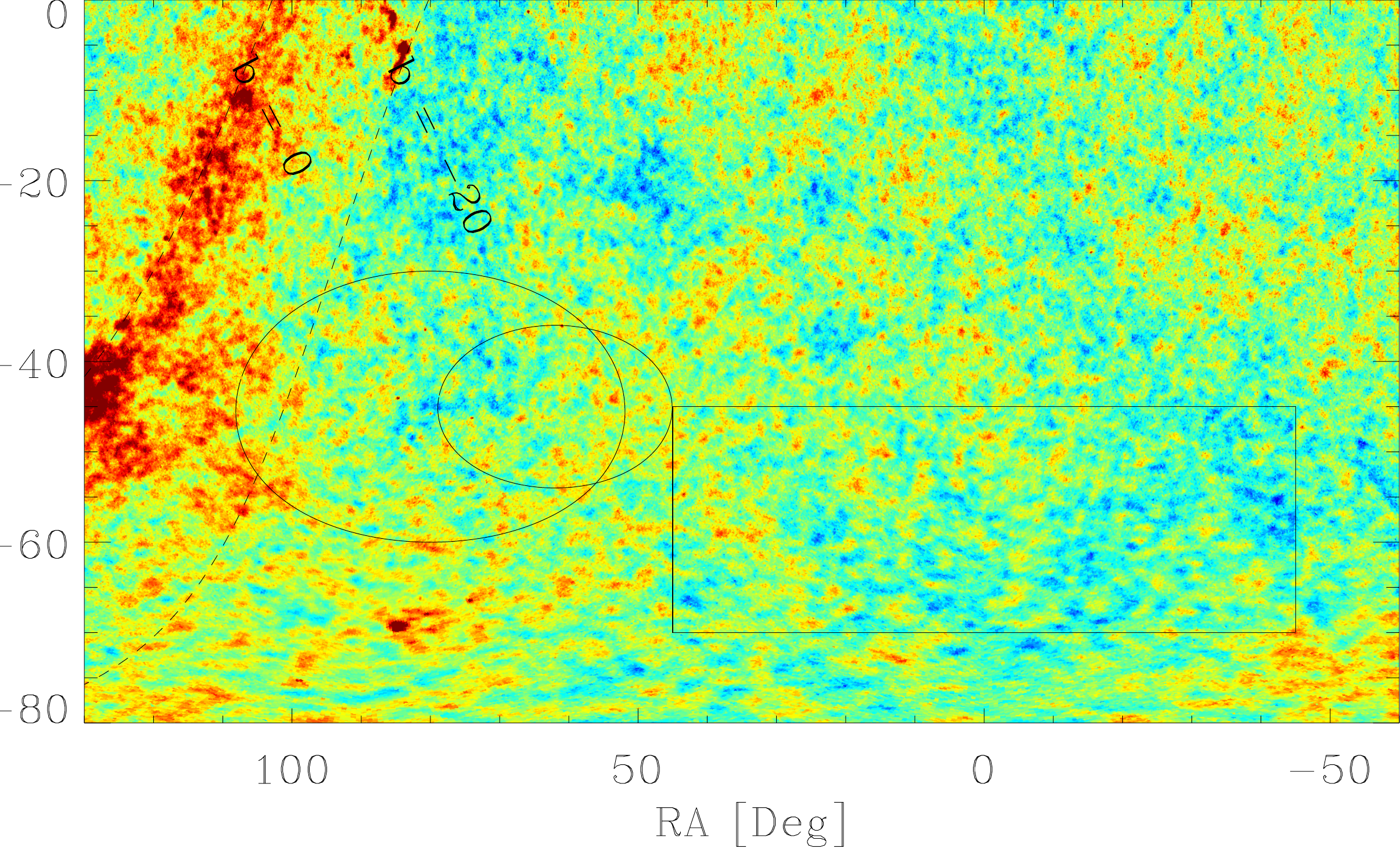}}
\scalebox{0.28}{\includegraphics[bb=-50 -20 700 420]{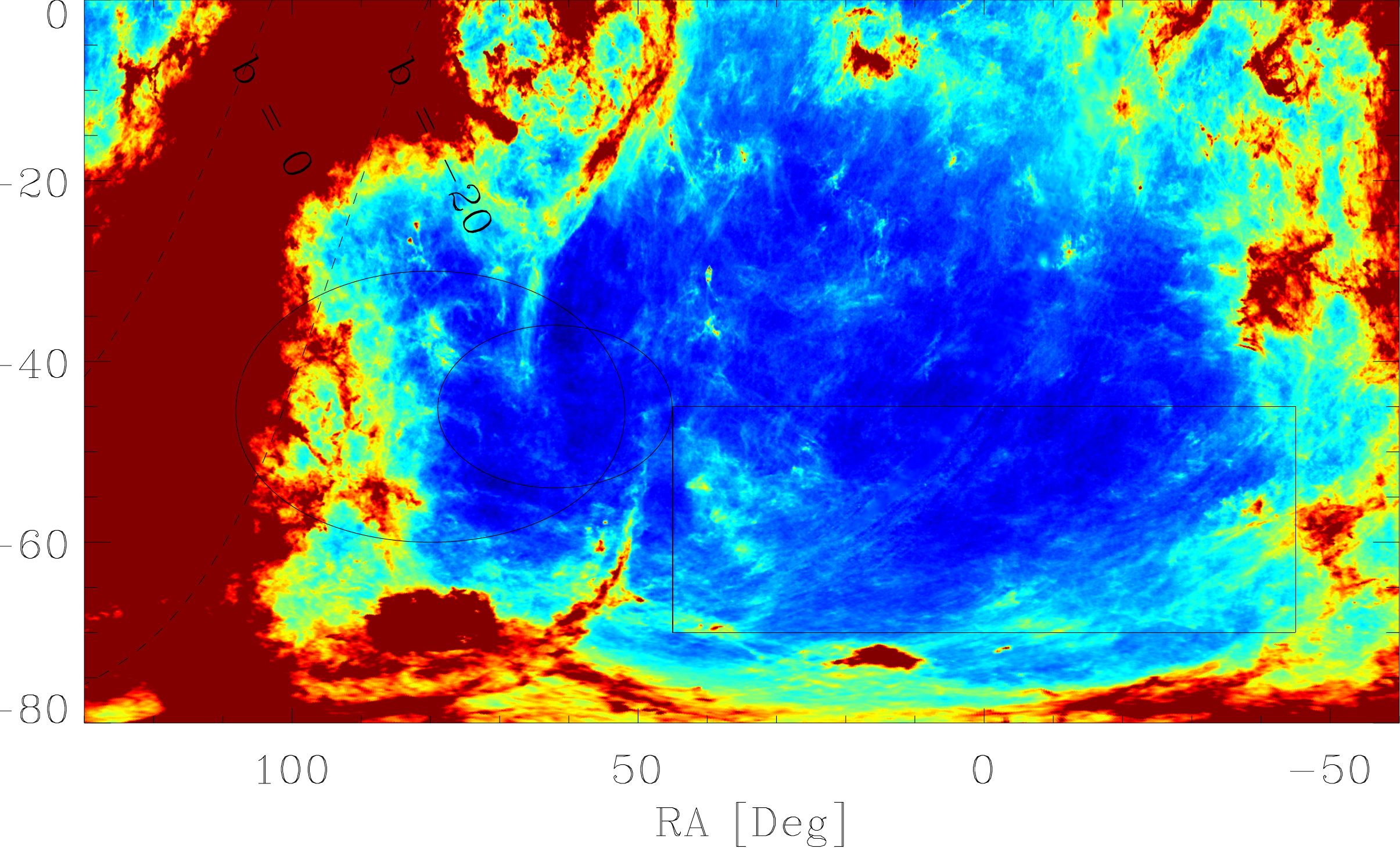}}
\end{center}
\caption{\small \emph{Upper left:} The sky coverage for the Antarctic LDB
  flight.  The integration time per pixel, for a single column of a \spider
  array, is shown in celestial coordinates.  Contours of galactic latitude are
  drawn for reference at $b=(0, -20)$.  Only a single column is shown to make
  the pattern of cross linking more evident.  The coverage of the full array
  will be smoothed in $\sim$RA by the $\sim$$8^\circ$ field of view.  Spider
  will cover about 8\% of the full sky, including the majority of the area in
  the Southern hemisphere that is not heavily contaminated by Galactic
  emission. The rectangular outline indicates the region of sky easily scanned
  from the geographic South Pole with low atmospheric loading and galactic foregrounds~\cite{Chiang:2009xsa}.  The large oval shows the
  CMB field observed by \boomerang \cite{MacTavish:2005yk,Ruhl:2002cz}, while the smaller
  oval shows the \ebex field~\cite{Grainger:2008spie}.  By virtue of its small,
  easily-shielded apertures, \spider can scan much closer to the sun (lower in
  RA) than \boomerang, and therefore cover more sky.  \emph{Upper right:} The
  seven year WMAP K-band data, a tracer of Galactic synchrotron emission\cite{Jarosik:2010iu}.  The
  bright point sources are mostly quasars at redshift $z\sim 0.5$ that will be
  used in \spider's pointing reconstruction.  \emph{Lower left:} The seven
  year WMAP W-band data.  The 90 GHz data are near the minimum of the total
  intensity of the Galactic foregrounds over much of the sky.  \emph{Lower
    right:} The IRAS $100~\mu$m map, showing the morphology of the thermal
  emission of Galactic dust\cite{Neugebauer:1984zz,Finkbeiner:1999aq}.}\label{fig:coverage}
\end{figure}

Figure \ref{fig:coverage} illustrates the expected sky coverage from this scan strategy and compares it to the known distribution of various galactic foregrounds.  \spider will map the cleanest area of the microwave sky, with a sky coverage much larger than that of many comparable instruments.


\section{FOREGROUNDS AND SYSTEMATICS}
\label{sec:syst}  

\spider is designed to limit the effect of galactic and atmospheric foregrounds upon the instrumental performance.  The locations and widths of the \spider bands are chosen to avoid prominent atmospheric spectral lines; in combination with the stratospheric vantage point provided by a balloon flight, this greatly reduces the atmospheric contributions of noise and excess loading.  \spider's 90 and 150 GHz bands target the frequencies where the CMB is brightest in comparison to galactic foregrounds, and are further optimized to avoiding the galactic CO line.  The 280 GHz band will map the polarized foreground emission of galactic dust; in combination with \planck maps of dust emission intensity, this will allow us to subtract any contribution from polarized dust emission with high accuracy.

Instrumental systematic effects present a huge challenge for the measurement of nanoKelvin polarization signals on the sky.  MacTavish \etal \cite{MacTavish:2008syst} have investigated the instrumental systematics of \spider using a simulation of the full data analysis pipeline.  This set of simulations determined a set of specifications on instrumental design and on in-flight or pre-flight knowledge of various instrumental parameters.  These initial simulations were focused upon a flight of \spider from Alice Springs, Australia, with greater sky coverage and a spinning-gondola scan strategy; detailed simulations of an Antarctic flight are still in progress.

\section{SPIDER PAYLOAD}
\label{sec:instrument}  

\begin{figure}[htbp]
   \centering
   \includegraphics[height=3in]{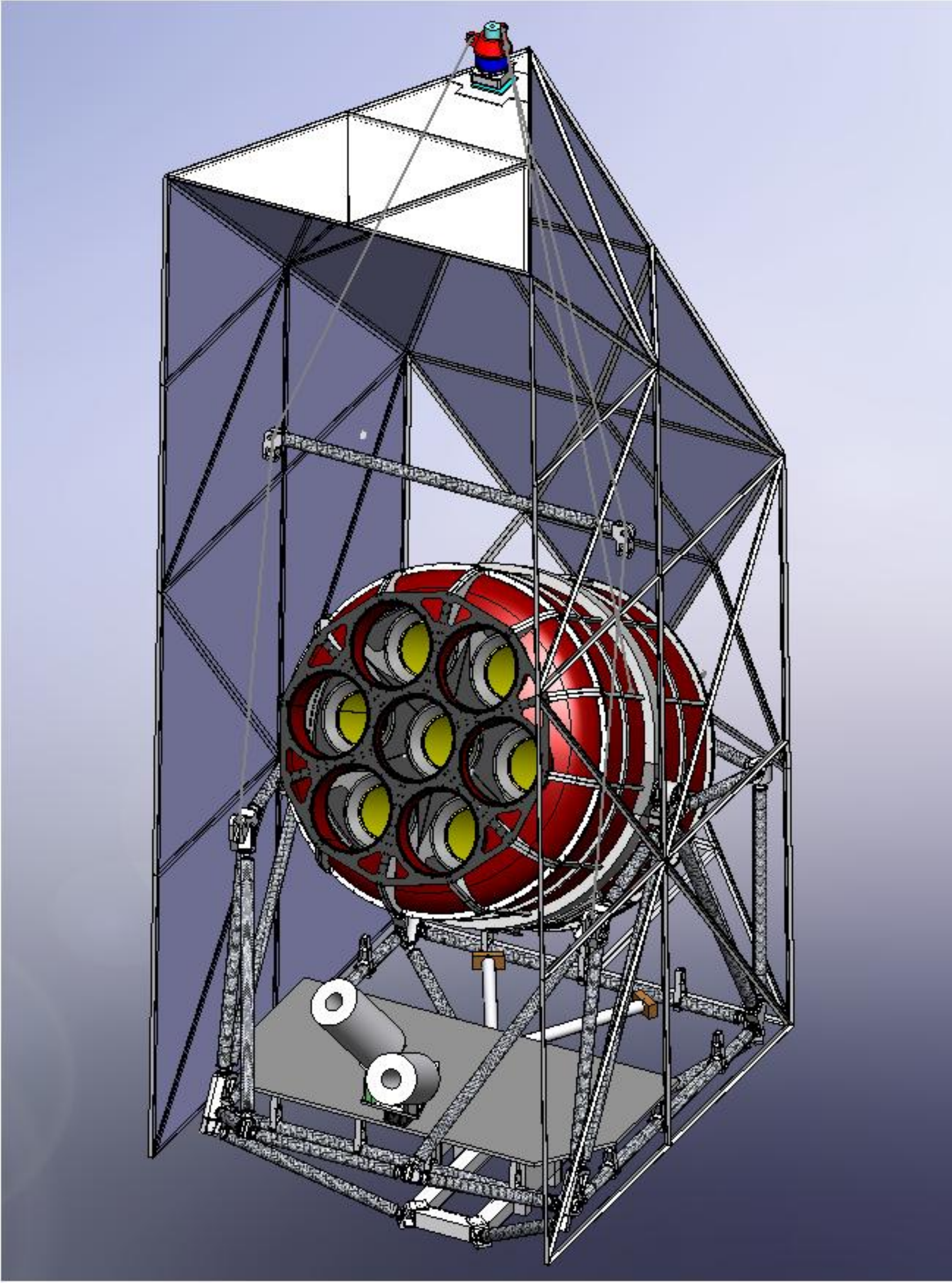} 
   \caption{An illustration of the \spider payload, showing the liquid helium cryostat and the surrounding carbon-fiber frame.  The bearing atop the frame allows the cryostat to scan in azimuth, while a linear actuator adjusts the bore sight elevation.  A sun shield protects the cryostat from solar heating.  The solar panels and electronics packages are not shown.}
   \label{fig:gondola}
\end{figure}

\subsection{Gondola}

The \spider payload, shown in Figure \ref{fig:gondola}, consists of six monochromatic refracting telescopes housed within a single liquid helium cryostat.  The gondola frame is derived from the successful \blast\cite{Pascale:2008blast} design.  The frame is constructed from carbon fiber tubes in order to maintain great strength at very low mass, allowing it to support a payload mass of $\sim$2000 kg under launch accelerations while itself weighing only $\sim$110 kg.  An azimuth reaction wheel allows the payload to operate in two major observing modes: back-and-forth scanning in azimuth or full azimuthal rotation; the former is most appropriate for daytime observation, the latter for operation at night.  The elevation of the inner frame is adjustable with a simple linear actuator similar to the \boomerang design.  The pointing and pointing reconstruction systems build on experience from the \boomerang\cite{Masi:2006inst} and \blast designs.  Pointing information is provided by two tracking star cameras, rate gyroscopes, differential GPS, and a sun sensor.  Extensive sun shielding and baffling, combined with the relatively small optical apertures of the \spider telescopes, allows the instrument to scan closer to the sun for increased sky coverage.

\subsection{Cryogenics}
The cryostat for the Spider instrument uses liquid helium-4 (LHe) to cool the instrument during its flight.  All six telescope inserts and a $\sim$1000 liter LHe tank are contained in a single outer vacuum vessel fabricated by Redstone Aerospace.  The primary LHe tank is maintained at 108 kPa ($\sim$1 atm) and has an expected minimum hold time of 25 days.  A smaller ($\sim$20 liter) superfluid LHe tank is continuously fed from the main tank through a capillary and controlled at a vapor pressure near 100 Pa to provide cooling power at $\sim$1.5 K.  Each focal plane is further cooled to 260 mK by its own closed-cycle $^3$He sorption refrigerator.  These sub-Kelvin coolers are recycled every 48-72 hours.  Further details of \spider's thermal architecture are described elsewhere in these proceedings\cite{Gudmundsson:2010spie}.

\subsection{Optical design}
The optical design of \spider is based on that of the successful {\it Robinson/\bicep} telescope \cite{Takahashi:2009vp,Aikin:2010spie}.  Each telescope is a monochromatic, telecentric refractor containing anti-reflection-coated polyethylene lenses cooled to 4 K.  The aperture field distribution of the primary is smoothly tapered with a 4 K Lyot stop to reduce detector background.  The beam of each pixel subtends 35 arc minutes (FWHM) on the sky.  The \spider telescope design is described further in a separate contribution to these proceedings\cite{Runyan:2008ip}.

A half-wave plate is mounted to the aperture of each telescope and cooled to 4 K.  The wave plate assembly consists of a single birefringent sapphire plate coated with a single layer of fused quartz on each side.  This assembly may be rotated in flight using a worm gear drive train, thus rotating the angle of polarization sensitivity on the sky while leaving the beams unchanged.  This allows a full measurement of the sky polarization using each individual detector, eliminating or reducing many potential systematic effects (\eg gain or pointing mismatches within a detector pair).  \spider's single-frequency telescopes simplify the wave plate design and implementation.  The optical performance of a prototype \spider wave plate assembly is described elsewhere in these proceedings\cite{Bryan:2010spie}.

The \spider optical system is constructed to take full advantage of the very low atmospheric background available at balloon altitudes by minimizing the stray light incident upon the detectors.  As described above, the principal optical elements - lenses, Lyot stop, and half-wave plate - are cooled to 4 K to reduce internal thermal radiation.  Detector beam sidelobes within the telescope tube terminate on an inner sleeve cooled to 2 K.  The focal plane is protected from out-of-band radiation by an extensive cooled filter stack containing infrared shaders, hot-pressed mesh filters, and a nylon filter.  In flight the \spider aperture will be covered by a thin-film polymer window with low emissivity; a thicker retractable HDPE window will be employed to withstand the greater atmospheric pressure on the ground.

\section{FOCAL PLANE}
\label{sec:detectors}  

\subsection{Focal plane}

The focal plane architecture of each \spider refractor is derived from that of the \bicep2 instrument, with extensive modifications to improve magnetic shielding\cite{Runyan:2010spie}.  Figure \ref{fig:FPU} shows a partially-assembled \spider focal plane unit.  Four Si bolometer arrays constitute the microwave-sensitive elements of each focal plane.  These detector tiles are supported at a distance of 1/4-wavelength above a superconducting Nb backshort plate, which defines an integrating cavity and provides magnetic shielding.   Flexible superconducting circuits connect the arrays to their associated SQUID amplifiers, which are housed within a magnetically-shielded box behind the focal plane.  The entire assembly is cooled to $\sim$300 mK by the $^3$He refrigerator.  The temperatures of each tile and of the copper support plate are read out using neutron-transmutation-doped (NTD) Ge thermistors.

\subsection{Bolometer arrays}

\begin{figure}[tbp]
   \centering
   \includegraphics[width=6.5in]{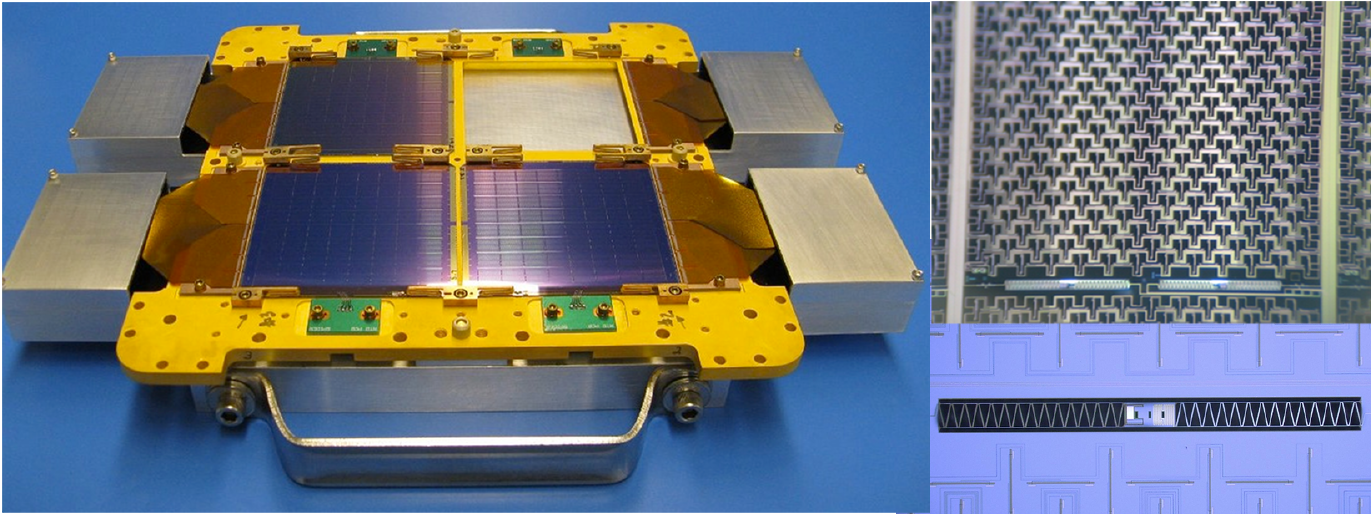} 
   \caption{{\it Left}: A partially assembled 150 GHz \spider test focal plane, with three of four detector tile slots filled.  The detector tiles each contain 64 spatial pixels and 128 TES bolometers.  The 16 SQUID multiplexer chips and their associated Nyquist inductor chips are encapsulated in four magnetically-shielded boxes behind the focal plane, connected to the detectors by flexible superconducting circuits visible to the right and left of the tiles; in this image the flexible circuits are covered by protective Al boxes which are not present in the final configuration.  The tiles are held in place by metal clips that maintain pressure despite thermal contraction. {\it Top right}: Closer view of a spatial pixel, showing phased array of slot antennas feeding two TESs at bottom. {\it Bottom right}: Close-up of a single TES assembly, showing the TES island and the SiN thermal isolation legs.  The meandered leg design allows for low thermal  conductance in a narrow geometry.}
   \label{fig:FPU}
\end{figure}

Each \spider refractor focuses light onto four large-format arrays of antenna-coupled bolometers \cite{Kuo:2008spie}, pictured in Figure \ref{fig:FPU}.  These arrays integrate beam and band defining elements with polarization-sensitive detectors into a single monolithic package, without a need for external feed horns.  Each array is a Si tile photolithographically patterned with an 8$\times$8 array of spatial pixels (6$\times$6 for 90 GHz arrays) with $\sim$$2F\lambda$ spacing.  A spatial pixel consists of two interleaved arrays of slot antennas, one each for the two perpendicular polarization components of incident light.  A summing tree of microstrip transmission lines link each array of slots to form a single antenna.  The slots are summed in-phase with one another, producing a diffraction-limited beam that couples to the optics with single-moded throughput.  This arrangement gives each pixel a  $\sim$13$^{\circ}$ FWHM beam, with a $\sim$25\% bandwidth (set by the slot dimensions) and a first sidelobe at -15 dB.    The output of each tree passes through an on-chip band-defining filter and terminates on a superconducting transition-edge sensor (TES) bolometer\cite{Irwin:2005tes}.  Systematic differences between the two polarization channels in each spatial pixel are limited by their shared optical path, the physical proximity of their bolometers, and their shared readout electronics chain.

The \spider bolometers are similar in design to those used in \bicep2\cite{Brevik:2010spie}, but have been optimized for the far lower optical background available at a suborbital platform.  The TES is a thin film of Ti ($T_c\approx500$ mK) on a silicon nitride membrane, thermally isolated from the tile substrate by narrow legs of silicon nitride.  Power received by the antenna is dissipated in a Au termination resistor on the island, which is maintained at the TES critical temperature by electrothermal feedback.  The \spider TES islands are designed to have somewhat lower island heat capacities and much lower thermal conductances to the tile than those used by \bicep2, as appropriate to the $\sim$5$\times$ lower background at float altitude.  The 150 GHz \spider detectors use a meandered leg design to achieve an extremely low thermal conductance ($G$=20 pW/K for a 150 GHz sensor) within a narrow geometry, the latter chosen to limit direct coupling of incident microwaves to the membrane.  These low conductances give the \spider arrays very low noise levels (see Table \ref{tbl:detectors}), but at the cost of relatively low TES saturation powers.  For beam characterization under laboratory loading conditions we make use of a secondary Al TES with much higher saturation power that is connected in series with the Ti TES.

\begin{figure}[tbp]
\begin{center}
\includegraphics[width=2.8in]{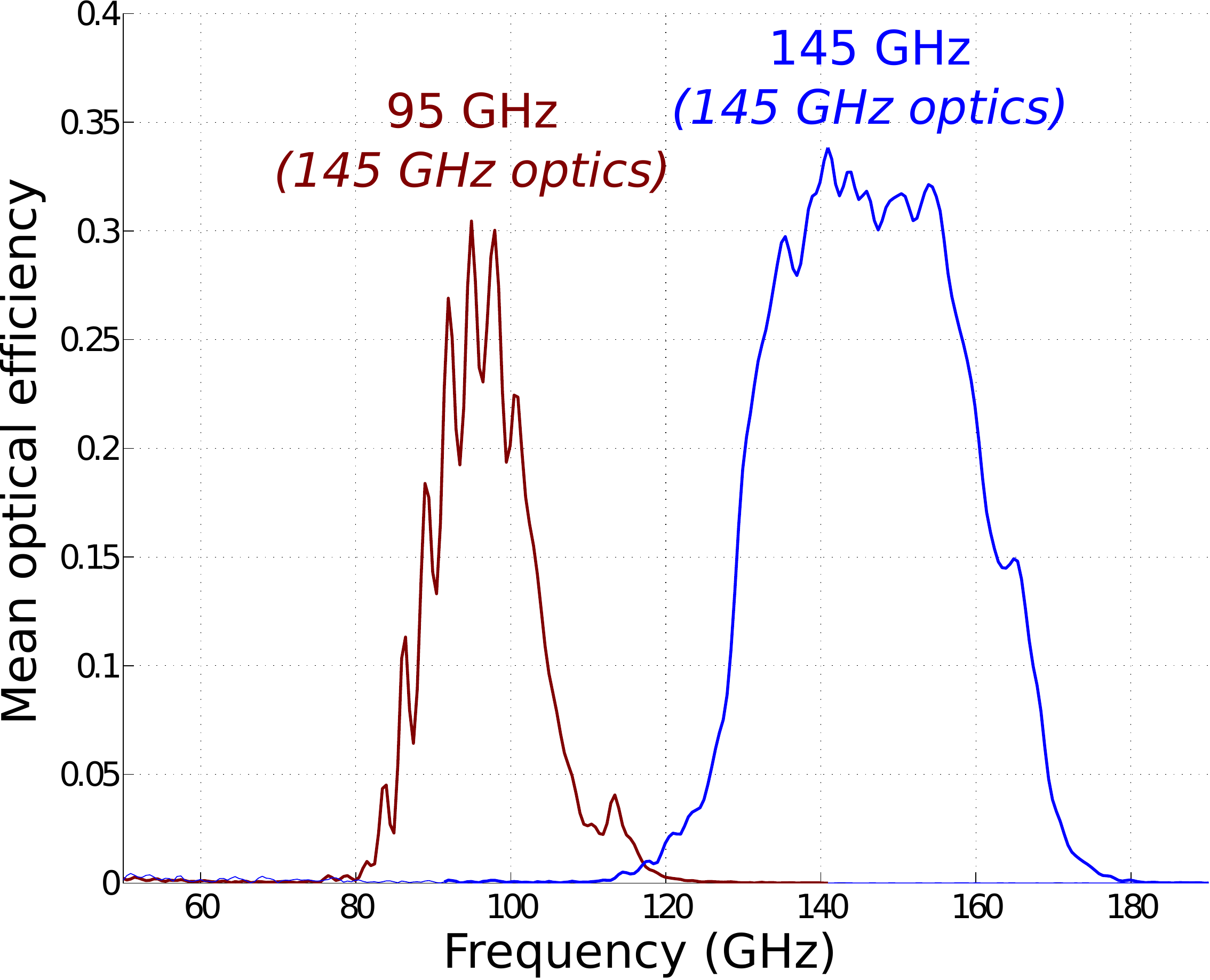}
\includegraphics[width=2.8in]{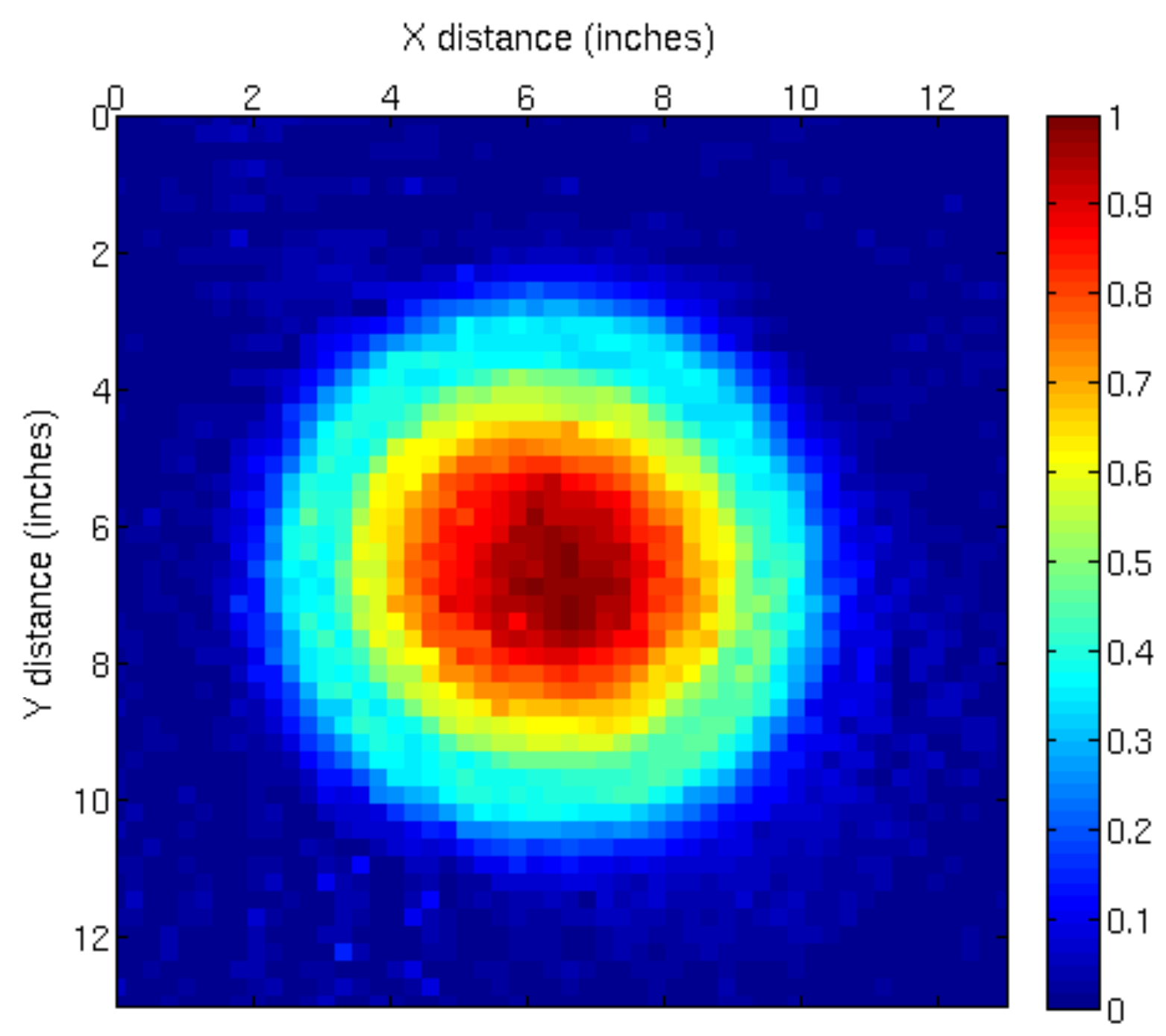}
\end{center}
\caption{\small \emph{Left:} Mean end-to-end frequency response of \spider 150 GHz and 96 GHz test detectors, measured through science-grade 150 GHz optics.  Note that the 96 GHz detector bands are distorted by the mismatched optics, introducing ripples in the passband and truncating the response before the upper band edge.  The band center of the science-grade detectors will be shifted from 96 GHz to $\sim$90 GHz to avoid background from O and CO lines.
\emph{Right:} Near-field beam map of a 150 GHz \spider detector, measured at the cryostat aperture through science-grade optics.
}\label{fig:det_performance}
\end{figure}

Figure \ref{fig:det_performance} shows results from optical characterizations of full detector arrays in the \spider test cryostat.  All measurements were taken in a 150 GHz \spider telescope, with a full set of optics appropriate for this frequency.  Optical efficiencies, band widths, and noise performance meet requirements for deployment.  We have not yet carried out a full characterization of the far-field response of the \spider beams, but results from preliminary laboratory tests and from measurements of the BICEP2 instrument in the field \cite{Aikin:2010spie} indicate that the arrays perform adequately for \spider's needs.  The \spider, \bicep2, and Keck collaborations continue to collaborate on improving the designs of these arrays.  Table \ref{tbl:detectors} lists the baseline frequency coverage and expected sensitivity of the full \spider instrument.

\begin{table*}[tbp]
\begin{center}
\begin{tabular}{|c|c|c|c|c|c|c|}
\hline
Band  &  & Beam & Number of &    & Single-Detector & Instrument \\
Center  & Bandwidth & FWHM & Spatial & Number of & Sensitivity & Sensitivity \\
(GHz) & (GHz) & (arcmin) & Pixels & Detectors & (\mukcmbrts)& (\mukcmbrts) 
    \\ \hline\hline
90 & 22 & 51  & ($2\times$)144 & ($2\times$) 288 & 110 (96) &  5 (4) \\
150 & 36 & 31  & ($2\times$) 256 & ($2\times$) 512 & 130 (80) &  4 (3) \\ 
280 & 67 & 17  & ($2\times$) 256 & ($2\times$) 512 & 470 (90) &  16 (3) \\ 
\hline
\end{tabular}
\end{center}
\caption{\small \spider observing bands, pixel and detector counts, and single-detector
and instrument sensitivities. The latter are quoted by dividing the
single-detector sensitivity by $\sqrt{N_\textrm{det}}$, assuming a detector yield of 85\%.  The quoted sensitivities at 90 and 150 GHz are typical of measurements made under flight-like loading conditions; the performance of the 280 GHz devices has not yet been measured under a representative load.  Sensitivities in parentheses are quoted in Rayleigh-Jeans units appropriate for dust emission.}
\label{tbl:detectors}
\end{table*}

\subsection{Readout system}

The hundreds of TESs making up each \spider focal plane are read out using a time-domain SQUID multiplexer system developed 
by NIST \cite{deKorte:2003mux}.  The changing current through each TES is detected by a corresponding first-stage SQUID.  These SQUIDs are switched on one at a time in rapid succession, so that each group of 33 SQUID channels (32 connected to bolometers plus one ``dark'' channel to monitor amplifier drifts) may be read out through a single set of wires to room temperature.  The output of each group of first-stage SQUIDs is amplified by a second-stage SQUID and a series SQUID array\cite{Huber:2001ssa} before reaching an ambient-temperature amplifier.  This arrangement greatly reduces the number of wires necessary to instrument these large-scale focal planes.  To maximize the effectiveness of pair-differencing in rejecting amplifier noise, the two polarizations of each spatial pixel are read out through adjacent first-stage SQUIDs in a common amplifier chain.

The TESs and SQUID multiplexer system are controlled and read out using the multi-channel electronics (MCE)\cite{Battistelli:2008mce}, an ambient-temperature electronics crate initially designed for the SCUBA-2 instrument \cite{Holland:2006spie}.  A single MCE crate maintains each focal plane, providing all necessary current and flux biases and linearizing the response of each SQUID channel using a digital feedback loop.  All communication with the gondola computers is through optical fiber.  A single crystal clock at 50 MHz synchronizes the timing of all six MCE crates on the gondola.  The MCE is highly programmable, allowing for flexible scripting and reconfiguration of science data acquisition and regular bolometer characterization.  In standard operation each TES is revisited every $\sim$66$\mu$s and data are recorded to solid-state disk at $\sim$400 Hz; other data acquisition modes are available for system characterization.

\subsection{Magnetic shielding}

By its nature, the \spider readout system is sensitive to the influence of external magnetic fields.  Each of the SQUIDs in the amplifier chain is a sensitive magnetometer, affected by tiny changes in the flux linking the SQUID loop.  The TESs themselves are also magnetically sensitive through the dependence of the superconducting film's critical temperature on magnetic field.  Each SQUID on the focal plane is wound as a second-degree gradiometer to reduce the influence of external fields, but changes in the focal plane's magnetic environment can nonetheless mimic sky signals.  Of particular concern is the motion of the gondola through the Earth's magnetic field, which could induce a scan-synchronous signal fixed in a terrestrial (rather than celestial or instrumental) coordinate system, corrupting low-$\ell$ modes of the map.  Magnetic shielding is thus a critical requirement for the \spider receivers.

The \spider telescope and focal plane are designed to shield the detectors and SQUIDs from external magnetic fields.  The \spider shield design incorporates several improvements upon the configuration used for BICEP2, developed with the aid of COMSOL simulations and extensive laboratory measurements.  Layers of high-$\mu$ material line each telescope tube and surround each focal plane assembly.  The Nb backshort further excludes magnetic fields from the vicinity of the focal plane.  The SQUID chips are further protected within a low-field enclosure behind the focal plane, encased inside several additional layers of high-$\mu$ and superconducting material.  No ferromagnetic materials were used in the construction of the telescope inserts or LHe cryostat.  \spider's magnetic shielding is described in greater detail elsewhere in these proceedings \cite{Runyan:2010spie}.

\section{CONCLUSIONS}
\label{sec:conclusions}  

\spider is unique as a powerful sub-orbital millimeter-wave polarimeter that can observe a large fraction of the sky.  By combining the massive mapping speed provided by large-format multiplexed TES arrays with an optical design optimized for polarimetry, it has the power needed to map the polarization of the CMB with unprecedented fidelity.  It's initial Antarctic flight will measure or powerfully constrain the B-mode signal from primordial gravitational waves, as well as making measurements of polarized foregrounds and of the lensing contribution to the CMB B-mode with lasting value.

\spider will also serve as a crucial pathfinder to test new technologies for EPIC/CMBpol, a proposed satellite mission to make definitive maps of the polarization of the millimeter-wave sky\cite{Bock:2009epic}.  The large-format detector array technology deployed in \spider is an ideal candidate for a space mission, allowing the rapid construction of kilopixel focal planes without the need for external feedhorn and filter hardware.  Following in the tradition of \boomerang and \blast, \spider will test these technologies under conditions directly applicable to a space mission: low optical background, a limited power budget, tight mass constraints, and autonomous operation.

\acknowledgments     
 
The \spider collaboration gratefully acknowledges the support of NASA (grant number NNX07AL64G), the Gordon and Betty Moore Foundation, and NSERC.  WCJ acknowledges the support of the Alfred P. Sloan Foundation.  JPF is partially supported by a Moore Postdoctoral Fellowship in Experimental Physics.

The \spider collaboration also extends special gratitude to Andrew E. Lange, who passed away on January 22.  Andrew was a leading light of the cosmology community and a driving force behind the \spider project.  He is greatly missed by his many collaborators, past and present.


\bibliography{Filippini_SPIE2010}   
\bibliographystyle{spiebib}   

\end{document}